\documentclass[%
 reprint, 
superscriptaddress,
colorlinks=true,linkcolor=blue,citecolor=blue,urlcolor=blue,
 amsmath,amssymb,
 aps, physrev, prm,
]{revtex4-2}
\usepackage[dvipsnames]{xcolor}
\usepackage{graphicx}
\usepackage{dcolumn}
\usepackage{bm}
\usepackage{booktabs} 
\usepackage[T1]{fontenc}
\usepackage[utf8]{inputenc}
\usepackage{siunitx}  
\DeclareSIUnit{\angstrom}{\text{\AA}}
\usepackage{multirow} 
\usepackage{subcaption}
\usepackage[percent]{overpic}
\usepackage{newtxtext,newtxmath} 

\usepackage[justification=raggedright,singlelinecheck=false]{caption}

\usepackage{orcidlink}
\usepackage{hyperref}
\begin{document}

\preprint{APS/PRM}

\title{\textbf{Comparative study of room temperature and quench condensed bismuth films: morphology and electronic characteristics} 
}%

\author{Yulia Kirina\,\orcidlink{https://orcid.org/0009-0006-5226-4982}}
\email{yulia@vt.edu}
\affiliation{Department of Materials Science and Engineering, Virginia Tech, Blacksburg, VA, USA}

\author{Prakash Sharma\,\orcidlink{https://orcid.org/0009-0002-1117-0956}}
\affiliation{Department of Materials Science and Engineering, Virginia Tech, Blacksburg, VA, USA}
\affiliation{Department of Physics, Virginia Tech, Blacksburg, VA, USA}

\author{Wyatt Thomas\,\orcidlink{https://https://orcid.org/0000-0002-1042-3053}}
\affiliation{Department of Materials Science and Engineering, Virginia Tech, Blacksburg, VA, USA}

\author{Tristan Anderson\,\orcidlink{https://orcid.org/0000-0001-8723-4026}}
\affiliation{Department of Physics, Virginia Tech, Blacksburg, VA, USA}

\author{Arya G. Pour\,\orcidlink{https://orcid.org/0009-0008-2813-6045}}
\affiliation{Department of Materials Science and Engineering, Virginia Tech, Blacksburg, VA, USA}
\affiliation{Department of Physics, Virginia Tech, Blacksburg, VA, USA}

\author{Victoria Soghomonian\,\orcidlink{https://orcid.org/0000-0001-8813-1053}}
\affiliation{Department of Physics, Virginia Tech, Blacksburg, VA, USA}

\author{Jean J. Heremans\,\orcidlink{https://orcid.org/0000-0002-6346-8597}}
 \email{heremans@vt.edu}
\affiliation{Department of Materials Science and Engineering, Virginia Tech, Blacksburg, VA, USA}
\affiliation{Department of Physics, Virginia Tech, Blacksburg, VA, USA}

\begin{abstract}
A comparison between properties of bismuth thin films deposited at substrate temperatures of 296 K (room temperature) and 77 K (quench condensed) is studied across epitaxial, amorphous, and van der Waals substrates. The experiments demonstrate changes in crystallinity, morphology, and electrical transport arising from the influence of substrate temperature. Moreover, the work highlights changes in grain size, roughness, X-ray diffraction peak intensities, and preferred orientation between the two deposition temperatures. The orientation of the films deposited at 77 K is preferentially (110), compared to (111) for films deposited at room temperature. Films grown at 77 K differ from those deposited at room temperature, exhibiting lower surface roughness but smaller grain size, which leads to increased electrical resistivity in quench condensed films. The decrease of substrate temperature during the deposition appears to induce slightly more strain in depositions on the amorphous and van der Waals substrates than on the epitaxial  substrates. Lastly, quench condensed films exhibit lower carrier mobility and lower carrier density compared to room temperature films. This study elucidates previously incompletely understood processes in bismuth deposition and raises new questions regarding growth on van der Waals surfaces.

\end{abstract}


\date{\today}

\maketitle


\section{\label{sec:intro}Introduction}
Thin film growth is central to microelectronics, sensors, nanoelectronics functionalities, and the exploration of new quantum material properties. Understanding the crystallographic structure, morphology, and electronic transport of thin films ranging from semiconductors to semimetals and topologically nontrivial materials is a step towards their utilization. One such material is bismuth. In bulk pure crystalline form it is a semimetal with equal hole and electron densities of about $3\times10^{23}$ $m^{-3}$ for $T \lesssim 30$ K. The densities at $T= 300$ K are higher, but still equal, around $2.5\times10^{24 }$ $m^{-3 }$ ~\cite{michenaud1972electron, hartman1969temperature}. Elemental bismuth harbors many unique physics phenomena: high spin-orbit interaction at its interfaces, low compensated carrier density and small effective masses all have implications for fundamental physics and spintronics~\cite{4,6,13}. Recent findings also include a possible non-linear Hall effect~\cite{7}, an anomalous Hall effect~\cite{7.2,8} and a two-dimensional topological insulator nature in ultrathin Bi films~\cite{3,5,9,14}. Bi is known to exhibit a higher superconducting transition temperature when deposited as a thin film at cryogenic substrate temperatures than in its bulk crystalline form~\cite{10,11,12,15,16,17}. 
The structure zone model for thin film growth \cite{18} suggests growth characteristics for specified temperature zones in terms of homologous temperature $T\textsubscript{h} = T / T\textsubscript{m}$ where $T\textsubscript{m}$ is melting temperature of the deposited material and $T$ is substrate temperature. When deposition occurs in Zone 1 (defined by $T\textsubscript{h} < 0.3$), limited surface mobility leads to reduced grain size due to insufficient energy for nucleation, often producing porous, tapered crystallites~\cite{18}. Bi, when deposited in Zone 1 conditions is often referred to as amorphous~\cite{10, 12}, and shows a different crystal structure than films obtained under room temperature (RT) deposition. Films deposited using the quench condensed (QC) method have been observed to undergo a superconducting transition around 5 K, unlike bulk bismuth which undergoes the transition at 0.53 mK ~\cite{5, 18}. Yet, full understanding of what happens to the structure as the temperature of the substrate decreases is unclear. Hitherto QC-Bi was only obtained by depositions below 25 K and only been compared to films deposited at RT or higher~\cite{8,10, 19, 20}. By investigating deposition between 25 K and RT a better understanding of the evolution of Bi growth can be obtained across temperatures. This work will look at Bi growth on three different substrates: amorphous SiO$_2$, epitaxial Al$_2$O$_3$(0001) and van der Waals mica. Each substrate has a different reason for being chosen; mica is known to be a layered material without dangling bonds that is used to achieve high quality growth through van der Waals epitaxy~\cite{21, 22, 23, 23.5}. Al$_2$O$_3$(0001) is a single crystal anisotropic material, with 4.6\% lattice mismatch with Bi(111) encouraging epitaxial growth of the Bi films~\cite{24, 27}. Lastly, SiO$_2$ is an amorphous substrate, which typically does not improve film growth.  SiO$_2$ will serve as a baseline to distinguish substrate dependent features from non-substrate-dependent ones. Crystallinity and morphology are analyzed and key differences are highlighted, between deposition temperatures and substrates. In addition, magnetotransport and sheet resistance are measured across all substrates to understand key changes in electronic properties between films.

\section{\label{sec:setup}methods}
Bi films of 100 nm thickness were grown on $(15\pm10) \times 10 ^{-2} \Omega\cdot\ m$ B-doped Si(100) wafer with a  \qty{3000}{\angstrom}  thick wet oxide of SiO$_2$, on Al$_2$O$_3$(0001) and on muscovite mica. After loading the substrates, the chamber was baked for 24 hours. For QC-Bi films, the substrate holder was cooled to 77 K by being in contact with the liquid nitrogen reservoir. The substrate holder was kept at constant temperature during the deposition. Pressure during deposition was $10^{-8}$ Torr. High purity (99.999\%) Bi was thermally evaporated from a tungsten boat, which is directly below the substrates. The deposition rate was kept constants for all samples, $1.2~ \si{\angstrom\per\second}$ . The film thickness was determined using a crystal monitor (Maxtek TM-100). 
Crystallinity of the Bi films was determined using a Panalytical X-ray diffractometer (XRD) system at X-Ray wavelength of 1.54 Å, scan between 15 and 75 degrees with step size of 0.0131 degrees. Morphology was determined using a Bruker atomic force microscope (AFM) with AC160 tips and AC Air topography mode, which is a no-contact mode. The scan size of all the micrographs is 5 by 5 $\mu\mathrm{m}$, number of points is 512, and scan rate is 1 Hz. Analysis of AFM data was performed using the Gwyddion software. Magnetoresistance and Hall resistance measurements were performed in the van der Pauw configuration. Six to twelve Ohmic contacts using Woods metal were attached to films making multiple four-point measurement segments. The magnetic field (B) was applied perpendicular to the film surface over the range of –1.4 T to 1.4 T at 296 K and 4.1 K. Sheet resistance (two-dimensional resistivity, denoted $R_{\square}$) was calculated from van der Pauw results $R_a$ and $R_b$. The $R_a$ and $R_b$ values are extracted from the data at B = 0 and are used in the expression \[
R_{\square} = \frac{\pi}{ln 2} \frac{(R_a + R_b)}{2}  f
\] where $f = 1 $ if $R_a=R_b$.

\section{Results and Discussion}
\subsection{\label{sec:structure}Morphology}
\begin{figure}[t]
    \centering
    \setlength{\abovecaptionskip}{-10pt}
    \setlength{\belowcaptionskip}{-10pt}

    \resizebox{\columnwidth}{!}{%
    \begin{tabular}{@{}c@{\hspace{0.5mm}}c@{}}
        \begin{subfigure}[t]{0.49\columnwidth}
            \includegraphics[width=\linewidth,trim=12 12 12 12,clip]{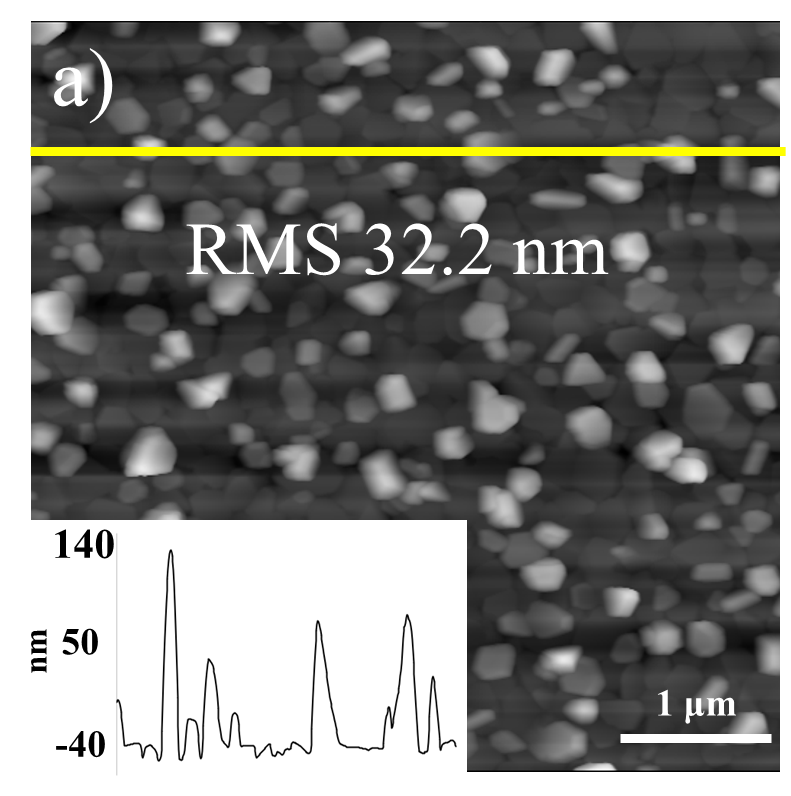}
            \phantomsubcaption\label{fig:sample_a}
        \end{subfigure} &
        \begin{subfigure}[t]{0.49\columnwidth}
            \includegraphics[width=\linewidth,trim=12 12 12 12,clip]{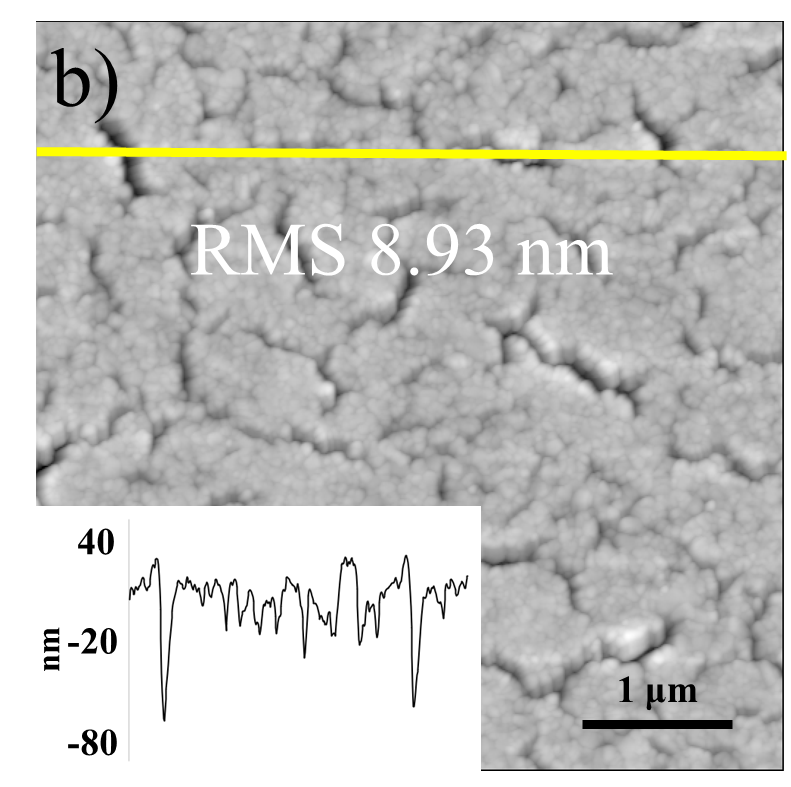}
            \phantomsubcaption\label{fig:sample_b}
        \end{subfigure} \\[-3mm]
\vspace{-5pt}
        \begin{subfigure}[t]{0.49\columnwidth}
            \includegraphics[width=\linewidth,trim=12 12 12 12,clip]{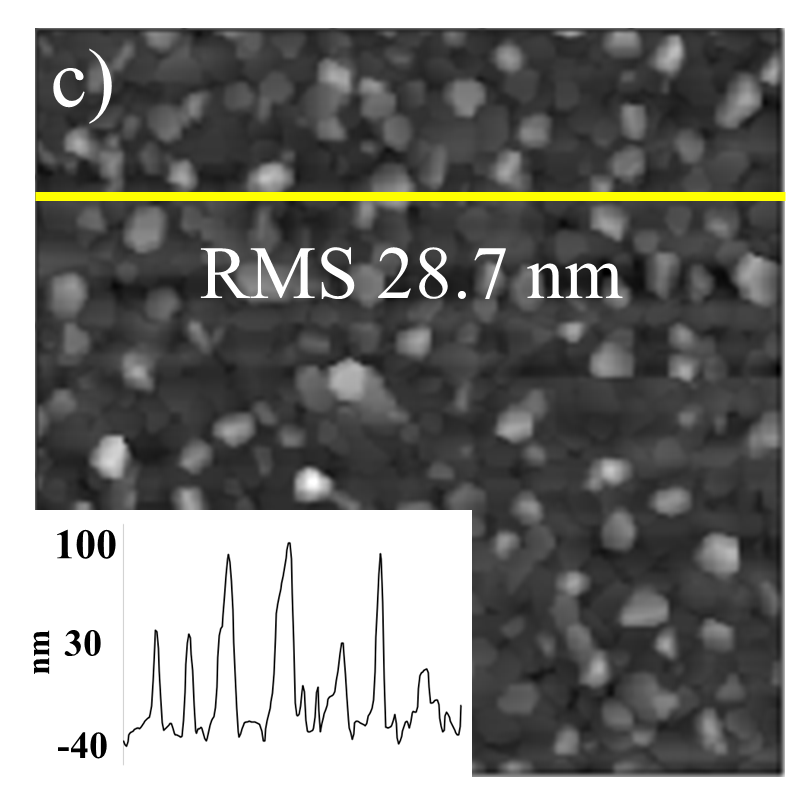}
            \phantomsubcaption\label{fig:sample_c}
        \end{subfigure} &
        \begin{subfigure}[t]{0.49\columnwidth}
            \includegraphics[width=\linewidth,trim=12 12 12 12,clip]{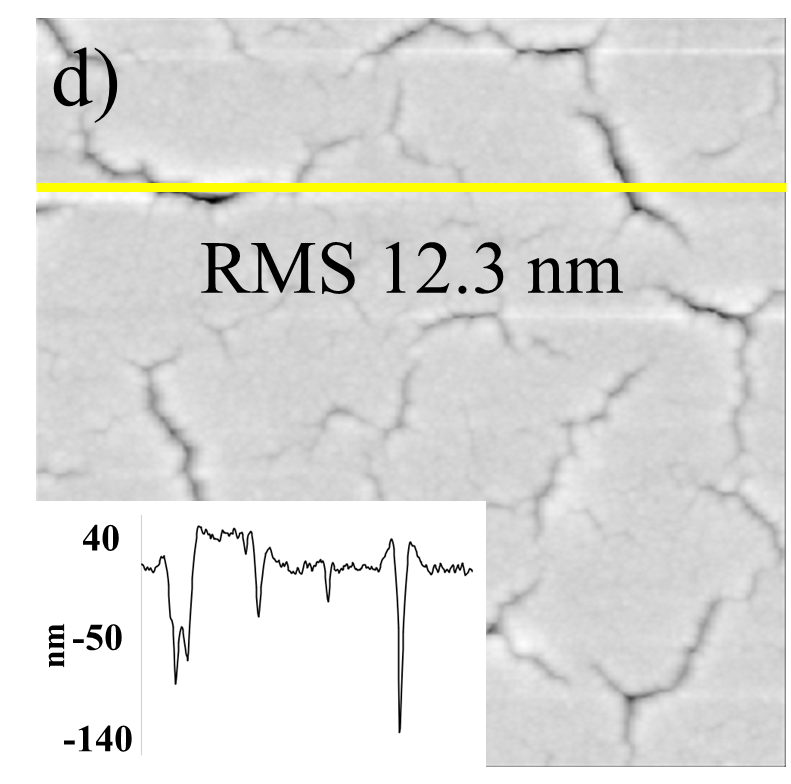}
            \phantomsubcaption\label{fig:sample_d}
        \end{subfigure} \\[-1mm]
\vspace{-5pt}
        \begin{subfigure}[t]{0.49\columnwidth}
            \includegraphics[width=\linewidth,trim=12 12 12 12,clip]{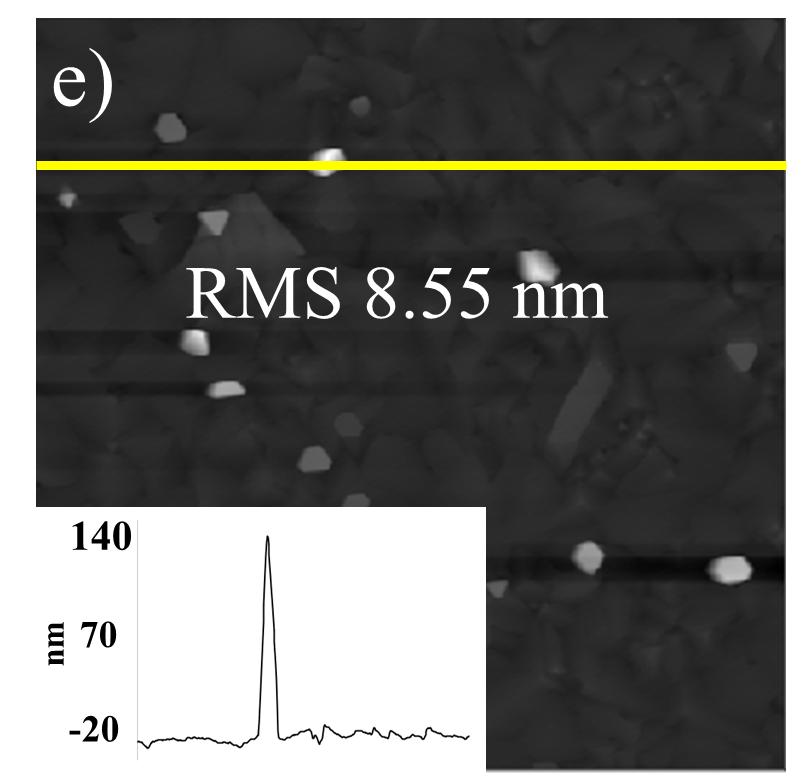}
            \phantomsubcaption\label{fig:sample_e}
        \end{subfigure} &
        \begin{subfigure}[t]{0.49\columnwidth}
            \includegraphics[width=\linewidth,trim=12 12 12 12,clip]{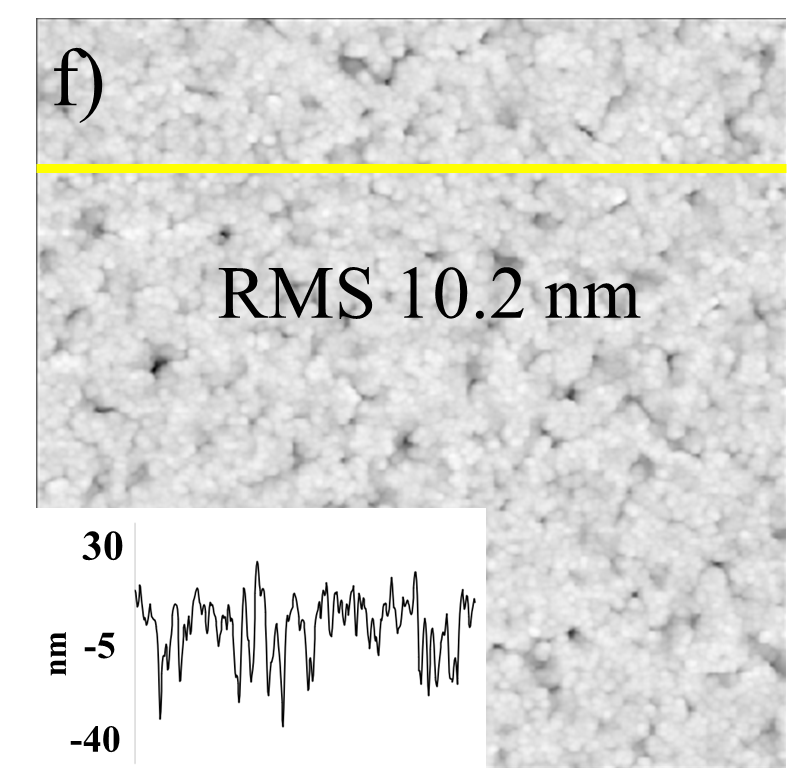}
            \phantomsubcaption\label{fig:sample_f}
        \end{subfigure}
    \end{tabular}%
    }
\vspace{10pt}
    \caption{AFM of Bi films: (a) $T_s=296$ K on Al$_2$O$_3$(0001); (b) $T_s=77$ K on Al$_2$O$_3$(0001); (c) $T_s=296$ K on SiO$_2$; (d) $T_s=77$ K on SiO$_2$; (e) $T_s=296$ K on mica; (f) $T_s=77$ K on mica. Also indicated for each panel are RMS roughness and a line scan of height along the indicated lines. }
    \label{fig:AFM}
\end{figure}

Figure~\ref{fig:AFM} shows typical AFM images of Bi films deposited on different substrates. Panels (\ref{fig:sample_a},~\ref{fig:sample_b}), (\ref{fig:sample_c},~\ref{fig:sample_d}), and (\ref{fig:sample_e},~\ref{fig:sample_f}) correspond to Al$_2$O$_3$(0001), SiO$_2$, and mica substrates, respectively, with substrate temperatures $T_s = 296$ K (\ref{fig:sample_a},~\ref{fig:sample_c},~\ref{fig:sample_e}) and 77 K (\ref{fig:sample_b},~\ref{fig:sample_d},~\ref{fig:sample_f}).

Root-mean-square roughness $R_q$ is measured from AFM images are shown in Fig.~{\ref{fig:AFM}}. Surface profiles overlaid on images of Fig.~{\ref{fig:AFM}} are obtained along the solid yellow lines. The profiles reveal that the surfaces of QC-Bi films on both Al$_2$O$_3$(0001) and SiO$_2$ (Fig. \ref{fig:sample_b}, and~\ref{fig:sample_d}) to be relatively smooth, with cracks on Al$_2$O$_3$(0001) being around 60 nm deep and 1 $\mu m$ long and on SiO$_2$ 100 nm deep and 1 $\mu m$ long. In contrast, cracks on QC/mica are not visible and the average depth is only about 30 nm. The RT-Bi films exhibit 3D morphologies with high aspect ratio are 20-120 nm tall and 100-200 nm wide, as seen in Figs.~\ref{fig:sample_a},~\ref{fig:sample_c},~\ref{fig:sample_e}. We find $R_q = 10.5\pm3.9$ nm for all QC-Bi films, irrespective of the substrate. In contrast, $R_q = 30.5\pm6.1$ nm as averaged for films on Al$_2$O$_3$(0001) and on SiO$_2$ substrate while it is $8.6\pm0.6$ nm as averaged for films of RT-Bi/mica, which is closer to the QC $R_q$ roughness value. Lower roughness on mica may be attributed to van der Waals epitaxy, where the substrate surface has no dangling bonds ~\cite{22, 23, 25, 26}. 
It is important to note that the morphology of QC-Bi films looks very similar to the films deposited at temperatures below 25 K ~\cite{24}. There are different general trends in the growth modes between RT-Bi and QC-Bi films. Room temperature growth of Bi in the literature is defined as Stranski-Krastanov, i.e. the film grows both in layers and in islands, reflected by the columns seen in Figs.~\ref{fig:sample_a},~\ref{fig:sample_c} and~\ref{fig:sample_e}. The nucleation in RT-Bi growth is known to be heterogeneous ~\cite{25, 27, 28}. QC-Bi growth appears to follow the random deposition model where atoms randomly stick on the substrate and little to no island formation occurs, as inferred from the considerably lower roughness~\cite{29}. The hypothesis is that the nucleation of QC-Bi film is homogeneous, and all sites nucleate simultaneously~\cite{27}. However, as the growth has not been observed \textit{in situ}, the present data does not allow a definite determination of the QC-Bi growth mode.

\subsection{Crytallinity}
\label{sec:2b}
\begin{figure}[t]
    \centering
    \setlength{\tabcolsep}{2pt}
    \hspace*{-15mm}
    \begin{tabular}{@{}c@{}c@{}}
    
\makebox[0.80\columnwidth][r]{%
    \includegraphics[height=4.2cm]{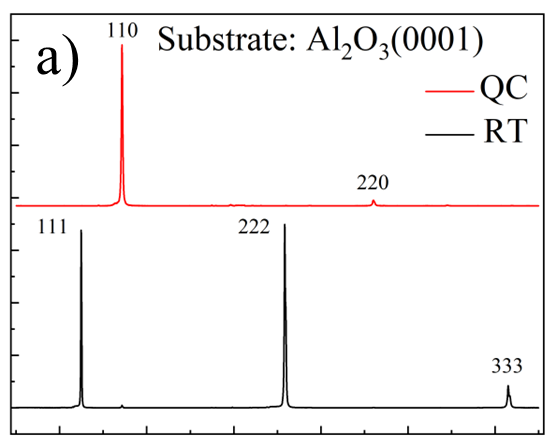}
}
\phantomsubcaption\label{fig:2a} \\[-3mm]

\makebox[0.80\columnwidth][r]{%
    \includegraphics[height=4.99cm]{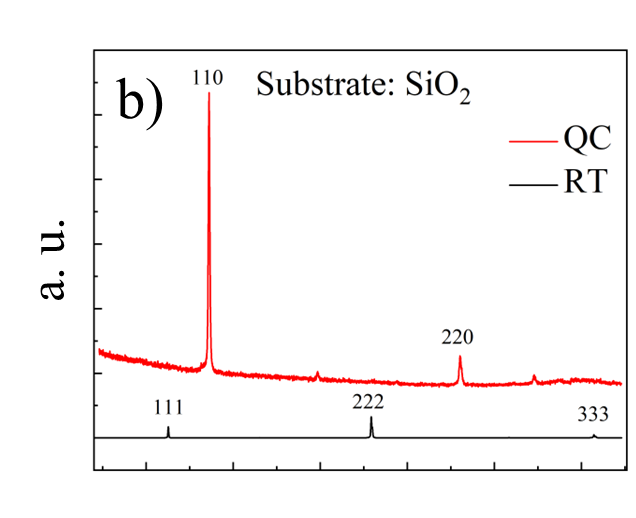}
}
\phantomsubcaption\label{fig:2b} \\[-3mm]

\makebox[0.80\columnwidth][r]{%
    \includegraphics[height=5.55cm]{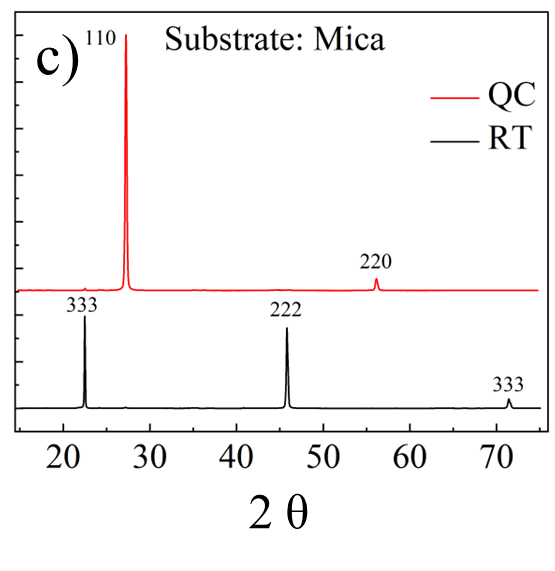}
}
\phantomsubcaption\label{fig:2c}
    \end{tabular}
    \vspace{-2mm}
     \caption{\label{fig:XRD} XRD data of QC and RT films on various substrates: a) $T_s$ = 296 K Bi film on Al$_2$O$_3$(0001), and $T_s$ = 77 K Bi film on Al$_2$O$_3$(0001);
b) $T_s$ = 296 K Bi film on SiO$_2$ and $T_s$ = 77 K Bi film on SiO$_2$; c) $T_s$ = 296 K Bi film on mica and $T_s$ = 77 K Bi film on mica. Substrate peaks have been subtracted for clarity. 
}
\end{figure}

We observe substrate temperature-dependent differences in the crystallinity of Bi films. Figure~\ref{fig:2a}~-~\ref{fig:2c} shows representative XRD scans from Bi films deposited at T$_s$ = 77 K and 296 K on a) Al$_2$O$_3$(0001), b) SiO$_2$, and c) mica. While multiple peaks are identified corresponding to the rhombohedral structure of Bi, we note that QC-Bi films have consistently fewer reflections compared to RT-Bi films~\cite{10}. QC-Bi films are primarily 110 textured and do not show clear 111 peaks while RT-Bi films have clear 110 peaks in addition to the 111 peaks. 
   
XRD peak width and AFM micrographs (Fig.~\ref{fig:AFM}) confirm that the films are all polycrystalline. We observe Bi 222 peak is the most intense in RT-Bi growth and Bi 110 is the most intense in QC-Bi growth. The peaks are suggestive of 110 texture in QC-Bi and 111 texture and RT-Bi on all three substrates~\cite{19}. 
To identify the influence, if any, of substrate structure on the Bi film texture, we compared the intensities of Bi 110 peaks in QC-Bi growth and 222 peaks in RT-Bi growth. To minimize any artifacts in peak intensities arising from the XRD measurement set up, we normalized the Bi film peaks with respect to the highest intensity peaks, Al$_2$O$_3$ 0006, Si 400, and mica 006 due to Al$_2$O$_3$(0001), SiO$_2$, and muscovite mica substrates respectively. We find that the Bi 110 (222) peak intensities are 61\% (70\%), 42\% (6\%), 54\% (42.8\%) of the Al$_2$O$_3$(0001) 0006, Si 400, and mica 006 peak intensities. These results suggest that 110 texture in QC-Bi is similar across all substrates, being the most pronounced on Al$_2$O$_3$(0001) and the least pronounced on SiO$_2$. Meanwhile, 222 texture exhibits a significantly larger variation in intensity across substrates but remains the most pronounced in RT-Bi films deposited on Al$_2$O$_3$(0001) and the least pronounced on SiO$_2$ substrates. 
Because the 110 peak is present in both RT-Bi and QC-Bi films, albeit with different intensities, it was used to quantify the peak shift and the associated strain between the two different $T_s$ growths. We do not observe a shift of the 110 peak for QC-Bi films on Al$_2$O$_3$(0001).
On SiO$_2$, there is a 0.003 degree shift of the 110 QC-Bi film peak position to lower angles relative to the 110 RT-Bi film peak position, indicating a small lattice expansion in QC-Bi film. On mica QC-Bi film, a shift of 0.004 degrees of the 110 peak to higher angles relative to the RT-Bi position of the 110 peak is observed, indicating a small lattice contraction in QC film. The XRD data also reveals information about comparative average grain sizes between QC-Bi and RT-Bi growth. Using the Scherrer equation, denoted below~\cite{30},  
 \[ d = \frac{K \lambda }{\beta cos(\theta)} \] 
 a statistical average of grain sizes can be calculated from XRD data using the FWHM (full width half max) of the peaks. Grain size was smaller in QC-Bi growth ranging from 37-42 nm in size, mica yielding the smallest and Al$_2$O$_3$(0001) yielding the largest. In RT-Bi growth the grain size ranged between 69-71 nm with Al$_2$O$_3$(0001) yielding the smallest, and SiO$_2$ yielding the largest. As the variation was not large, the grain size appears to be independent of substrate, but dependent on substrate temperature. In summary QC-Bi growth decreases grain size by around 30 nm and changes the preferred orientation to 110 from 111 when compared to RT-Bi.

\subsection{\label{sec:MR}Electronic Characteristics}
Magnetotransport measurements can reveal electronic properties of the films such as type of carriers, carrier densities and mobilities, as well as physical phenomena such as semimetal-to-semiconductor transitions, etc. This information can be extracted using sheet resistance (two-dimensional resistivity), magnetoresistance, and Hall resistance~\cite{26, 19, 31, 32}. Electrical measurements support morphological characterization and help with a more comprehensive understanding of the films.  
Sheet resistances for RT-Bi and QC-Bi films on Al$_2$O$_3$(0001), SiO$_2$ , and mica, were measured with a 10 $\mu A$  current, at 296 K, and 4.1 K, with data reported in Table~\ref{tab:table}. 
\begin{table}
\centering
\caption{\label{tab:table}Electrical properties at 296~K and 4.1~K for QC and RT bismuth films grown on SiO$_2$, Al$_2$O$_3$(0001), and mica.}
\begin{tabular}{l l
                cccc
               }
\toprule
\multirow{2}{*}{Substrate} & & \multicolumn{2}{c}{296~K} & \multicolumn{2}{c}{4.1~K} \\
\cmidrule(lr){3-4} \cmidrule(lr){5-6}
 & & \mbox{${R_{\square}}(\Omega)$} & \mbox{$n(10^{26}\mathrm{m^{-3}})$} 
     & \mbox{${R_{\square}}(\Omega)$} & \mbox{$n(10^{26}\mathrm{m^{-3}})$} \\
\midrule
\textbf{QC} & SiO$_2$        &  58.9 & 6.79 & 115 & 7.76 \\
            & Al$_2$O$_3$(0001) & 121 & 0.67 & 136 & 9.37 \\
            & Mica           & 22.3 & 1.25 & 114 & 1.05 \\
\midrule
\textbf{RT} & SiO$_2$        & 19.3 & 3.21 & 339 & 0.26 \\
            & Al$_2$O$_3$(0001) & 17.3 & 3.41 & 65.4 & 0.36 \\
            & Mica          & 3.42 & 7.70 & 13.0 & 0.81 \\
\bottomrule
\end{tabular}
\end{table}

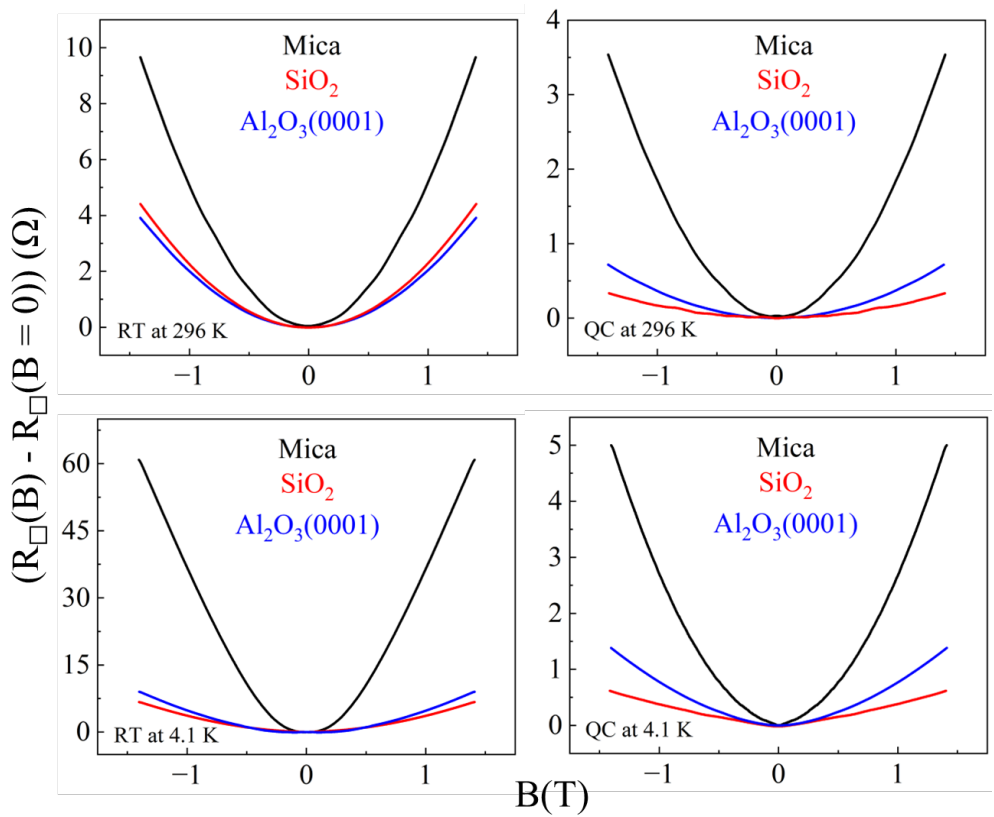
\begin{figure*}[!t]
    \includegraphics[scale =0.6]{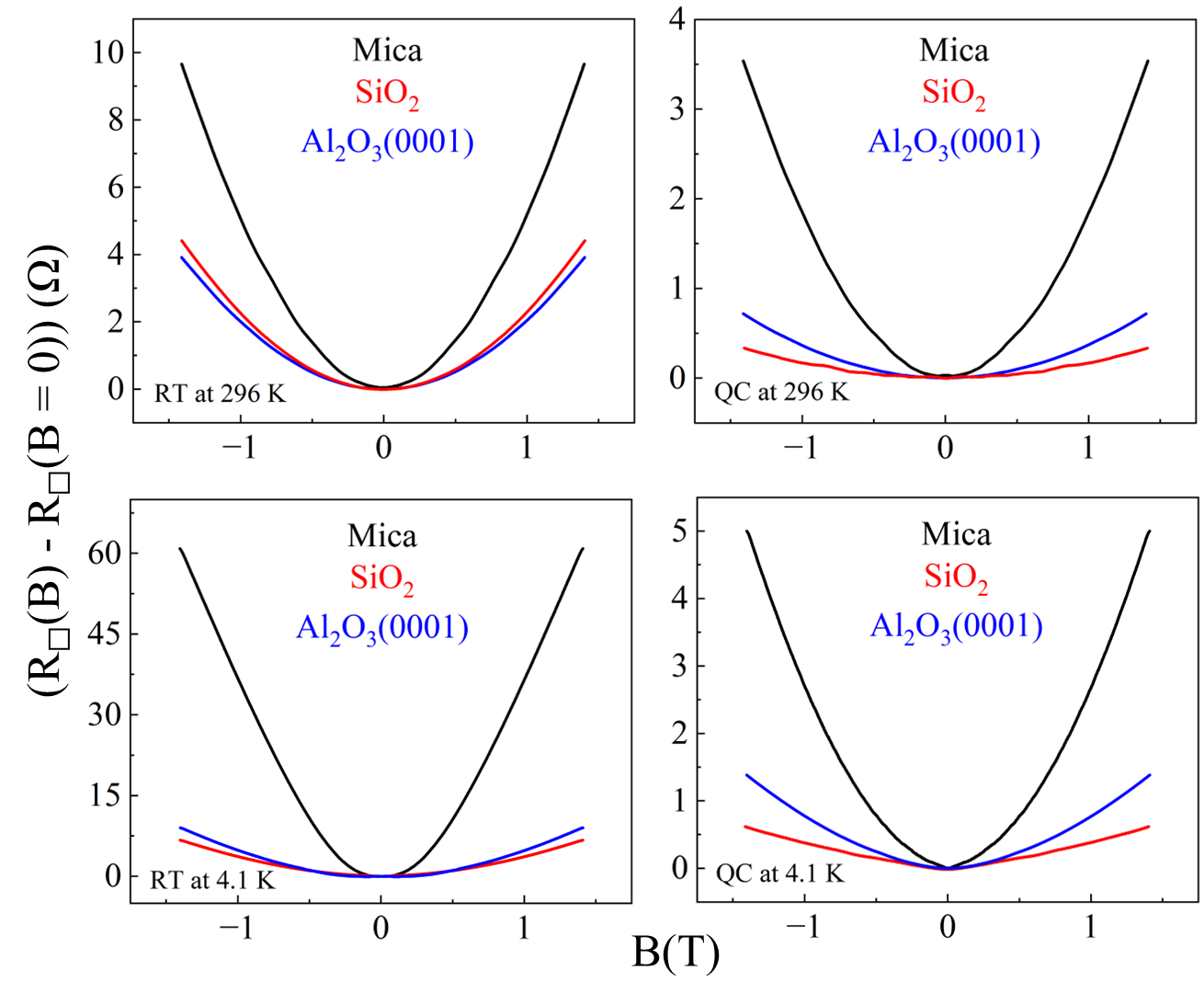}
    \setlength{\unitlength}{1pt}
    \begin{picture}(0,0)
        \put(115,410){\phantomsubcaption\label{fig:3a}}
        \put(365,410){\phantomsubcaption\label{fig:3b}}
        \put(115,170){\phantomsubcaption\label{fig:3c}}
        \put(365,170){\phantomsubcaption\label{fig:3d}}
    \end{picture}
    \caption{\label{fig:Rxx} Relative magnetoresistance vs magnetic field B of all Bi films, for: a) RT grown Bi films, at 296 K; b) QC grown Bi films, at 296 K;  c) RT grown Bi films, at 4.1 K; d) QC grown Bi films, at 4.1 K.	
}
\end{figure*} 
B is applied perpendicularly to the film plane. Figure \ref{fig:Rxx} compares relative magnetoresistance, i.e. with the value at B=0 subtracted, for QC-Bi and RT-Bi films on Al$_2$O$_3$(0001), SiO$_2$, and mica at 296 K and 4.1 K.  
Since all growths were of the same thickness of 100 nm, comparing sheet resistances of different films is equivalent to comparing their 3D resistivities, which in turn reflects the electronic properties irrespective of film thickness. Results show QC-Bi films exhibited higher sheet resistance compared to RT-Bi films across all substrates at 296 K measurements. The increase in sheet resistance is likely because QC-Bi films have smaller grains increasing the total number of grain boundaries, as mentioned in Sec.~\ref{sec:2b}. The cracks observed on the AFM micrographs also add to discontinuity of films likely causing an increase in sheet resistance~\cite{33, 34}. Between the three substrates, Bi films on mica demonstrated lower sheet resistance than films on Al$_2$O$_3$(0001) and SiO$_2$ , which is attributed to strain-free van der Waals epitaxial growth on mica ~\cite{21, 22, 23, 26}. Under RT-Bi growth, the order from highest to lowest sheet resistance comprises growth on SiO$_2$, Al$_2$O$_3$(0001) and mica. Meanwhile, in QC-Bi growth as measured at 296 K, growth on SiO$_2$ shows lower sheet resistance than on Al$_2$O$_3$(0001). To compare how sheet resistance changes between RT-Bi and QC-Bi growth we divide the sheet resistances of QC-Bi growths by sheet resistances of RT-Bi growths, always yielding a ratio $>$ 1. We obtain by comparing QC-Bi growth vs RT-Bi growth, the sheet resistance of films on mica increases by a factor of 6.5, of films on Al$_2$O$_3$(0001) increases by a factor of 6.9 and of films on SiO$_2$ increases by only a factor of 3.1. The observation that the ratio is the lowest in SiO$_2$ is potentially attributed to the finding that QC-Bi films on SiO$_2$ possess induced compressive strain, which decreases the energy bandgap in the 110 orientation, hence lowering their sheet resistance ~\cite{34}. 

Magnetoresistance measurements consisting of sheet resistance vs magnetic field (B) from -1.4 to 1.4 T, are represented in Figs. ~\ref{fig:3a}~–~\ref{fig:3d}. 
\begin{figure*}[!t]
    \includegraphics[scale =0.33]{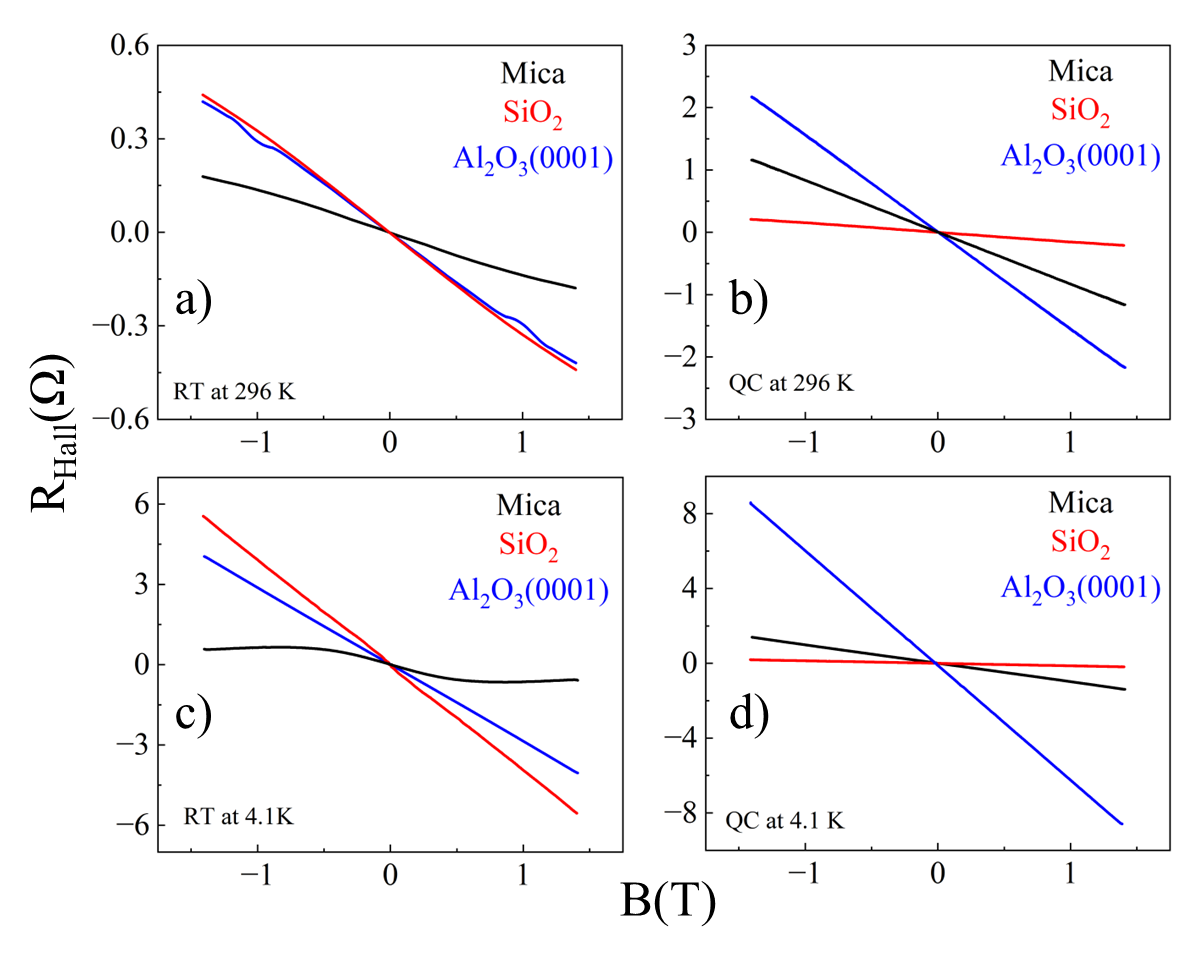}
    \setlength{\unitlength}{1pt}
    \begin{picture}(0,0)
        \put(115,410){\phantomsubcaption\label{fig:4a}}
        \put(365,410){\phantomsubcaption\label{fig:4b}}
        \put(115,170){\phantomsubcaption\label{fig:4c}}
        \put(365,170){\phantomsubcaption\label{fig:4d}}
    \end{picture}
    \caption{\label{fig:Hall} Hall resistance vs magnetic field B of all Bi films, for: a) RT grown Bi films, at 296 K; b) QC grown Bi films, at 296 K, where the mica film shows multicarrier behavior; c) RT grown Bi films, at 4.1 K, where the mica film shows multicarrier behavior; d) QC grown Bi films, at 4.1 K.
}
\end{figure*}
In Figs.~\ref{fig:4a}~-~\ref{fig:4d}~ the Hall (transverse) resistance is plotted as a function of B for QC-Bi and RT-Bi films on these substrates at the same temperatures. Multicarrier models of the magnetoresistance and the Hall resistance take into account the existence of multiple carrier types ~\cite{26, 31}. In Bi, the magnetoresistance and the Hall resistance vs B can be fitted to a 3-carrier model of magnetotransport, which includes the dominant carriers in high quality bulk Bi crystals, namely electrons and holes in the bulk of the film and surface state electrons ~\cite{26, 31}. 
In multicarrier analysis the nonlinearity in the Hall resistance vs B for RT-Bi on mica in Figs.~\ref{fig:4a}, ~\ref{fig:4c} indicates the presence of multiple carrier types, while the linearity for RT-Bi on Al$_2$O$_3$(0001) and SiO$_2$ indicates transport by a single dominant carrier type on these substrates. The linearity of the Hall resistance vs B for all QC-Bi films, on Al$_2$O$_3$(0001), SiO$_2$ and mica, indicates a single dominant carrier type in QC-Bi films as well. The quadratic magnetoresistance in Fig.~\ref{fig:Rxx} for all films can also be indicative of multiple carrier types but constitutes less strong evidence of multicarrier transport because magnetoresistance can also originate from sample geometry especially if carrier mobilities are high (geometrical magnetoresistance). We therefore take stronger magnetoresistance as a qualitative indication of higher carrier mobility, fewer defects leading to carrier scattering, and less degraded electronic properties relative to crystalline bulk Bi. Table~\ref{tab:table} reports carrier densities determined from Hall resistance of the films assuming single dominant carrier types (the carrier density values for RT-Bi on mica hence could be inaccurate but still indicate good approximations). 
In all QC-Bi films, and for RT-Bi on Al$_2$O$_3$(0001) and SiO$_2$ the sign of the slope of the Hall resistance vs B indicates transport by electrons, consistent with findings in ~\cite{33,34}. As shown by the nonlinear Hall resistance of RT-Bi on mica (Figs.~\ref{fig:4a}, ~\ref{fig:4c}), multicarrier transport asserts itself, indicating electronic properties more closely approaching crystalline bulk properties. Indeed in reverse, the existence of a single dominant carrier type in materials where we expect multiple carriers to contribute, is often suggestive of unintentional doping by defects produced during growth. The strongest magnetoresistance ~\cite{35} in Figs.~\ref{fig:3a}, ~\ref{fig:3c}, as shown by RT-Bi on mica, is indicative of a higher carrier mobility in these films than in any other films discussed, and hence of fewer defects leading to carrier scattering, consistent with the analysis of the nonlinear Hall resistance. Notably, even QC-Bi on mica shows high magnetoresistance (Figs.~\ref{fig:3b}, ~\ref{fig:3d}) indicating Bi on mica consistently results in high quality growth and commensurate electronic properties. When analyzing the substrate influence via the magnetoresistances in Fig.~\ref{fig:Rxx}, growth on mica shows highest mobility and SiO$_2$ shows the lowest, regardless of measurement temperature or growth type. RT-Bi growth qualitatively results in stronger magnetoresistance indicating it results in films with fewer defects and with higher carrier mobility. 
Carrier mobility is lowered by scattering on Coulombic lattice defects, which does not show a notable dependence on temperature, and by scattering on phonons, which strongly diminishes at lower temperature. A mobility tending to remain constant over temperature indicates dominant defect scattering and hence a more disordered film. Taking stronger magnetoresistance as indicative of higher carrier mobility, the relative increase in magnetoresistance upon cooling from 296 K to 4.1 K (Fig.~\ref{fig:Hall}) reflects the role of defect scattering. A larger increase in magnetoresistance with decreasing temperature indicates reduced defect scattering and, consequently, lower disorder. Figure~\ref{fig:Hall} reveals RT-Bi films on mica show the strongest increase in magnetoresistance when cooled from 296 K to 4.1 K, and hence likely are the least disordered films. QC-Bi films on mica by this trend also show relatively low disorder. Both RT and QC-Bi films on SiO$_2$ reveal a higher disorder among all films, while films on Al$_2$O$_3$(0001) show disorder intermediate between films on mica and SiO$_2$. 
Table~\ref{tab:table} shows the density of carriers for RT-Bi films consistently shows higher values at 296 K compared to 4.1 K, as expected for Bi due carrier freeze-out resulting from a semimetal bandstructure with small band overlaps and small bandgaps~\cite{26}. In QC-Bi films an increase in temperature is not consistently accompanied by a notable increase in carrier density. The reason for the unusual dependence on temperature of carrier density in QC-Bi films is presently unknown, but correlates with the unusual growth conditions. Explanations could lie in an interplay play of complex band structure of both Bi 110 and Bi 111 textures, mixed with different types of strain induced by the substrates and quantum confinement, leading to different results~\cite{10, 35, 36, 37, 38}.

\section{Conclusion}
In conclusion, we present a comparison of physical properties between QC-Bi deposition at 77 K and RT-Bi deposition at 296 K on three different substrates, Al$_2$O$_3$(0001), SiO$_2$, and mica. AFM micrographs showed QC-Bi films are smoother with crack features on Al$_2$O$_3$(0001) and SiO$_2$, and with pore features on mica. RT-Bi films exhibit 3D morphologies with high aspect ratio. Roughness analysis reveals that Bi films grown on mica exhibit significantly lower surface roughness than those on the other substrates. XRD scans showed a change in preferred orientation from Bi 111 on RT-Bi films to Bi 110 at 77 K. The grain sizes of QC-Bi films across all substrates are about half of RT-Bi grain sizes, calculated from XRD data. Included is an analysis of the electronic transport properties at 4.1 K and 296 K of magnetoresistance, Hall resistance and sheet resistance. QC-Bi films had higher sheet resistance, weaker magnetoresistance and lower carrier density than RT-Bi films. Across substrates, films grown on mica exhibit lower sheet resistance and stronger magnetoresistance under both QC and RT conditions, consistent with higher crystalline quality associated with van der Waals epitaxy. These results highlight the strong interplay between deposition temperature, substrate, and microstructure in determining the electronic properties of Bi thin films.

\begin{acknowledgments}
 We thank Prof. Suneel Kumar Kodambaka for helpful feedback and supervision. We acknowledge the support from the National Science Foundation under Award No. 2245008. Lastly, we thank and appreciate the facilities provided by the Institute for Critical Technology and Applied Science and their staff for guidance on sample characterization.  
  
\end{acknowledgments}



\nocite{*}

\bibliography{Reference}

@article{michenaud1972electron,
doi = {10.1088/0022-3719/5/21/011},
url = {https://doi.org/10.1088/0022-3719/5/21/011},
year = {1972},
month = {oct},
publisher = {},
volume = {5},
number = {21},
pages = {3061},
author = {J - P Michenaud and J -P Issi},
title = {Electron and hole transport in bismuth},
journal = {Journal of Physics C: Solid State Physics},
}

@article{hartman1969temperature,
  title = {Temperature Dependence of the Low-Field Galvanomagnetic Coefficients of Bismuth},
  author = {Hartman, Robert},
  journal = {Physics Review},
  volume = {181},
  issue = {3},
  pages = {1070--1086},
  numpages = {0},
  year = {1969},
  month = {May},
  publisher = {American Physical Society},
  doi = {10.1103/PhysRev.181.1070},
  url = {https://link.aps.org/doi/10.1103/PhysRev.181.1070}
}

@article{3,
  doi = {10.1088/1367-2630/16/5/055004},
url = {https://doi.org/10.1088/1367-2630/16/5/055004},
year = {2014},
month = {may},
publisher = {IOP Publishing},
volume = {16},
number = {5},
pages = {055004},
author = {Takayama, A and Sato, T and Souma, S and Takahashi, T},
title = {Rashba effect in antimony and bismuth studied by spin-resolved ARPES},
journal = {New Journal of Physics},
}

@article{4,
  title = {Densities of charge-carriers in bismuth},
journal = {Physics Letters A},
volume = {26},
number = {10},
pages = {490-491},
year = {1968},
issn = {0375-9601},
doi = {https://doi.org/10.1016/0375-9601(68)90807-4},
url = {https://www.sciencedirect.com/science/article/pii/0375960168908074},
author = {J.M. {Noothoven Van Goor}}
}

@article{5,
   title = {Bi(111) Thin Film with Insulating Interior but Metallic Surfaces},
  author = {Xiao, Shunhao and Wei, Dahai and Jin, Xiaofeng},
  journal = {Physics Review Letters},
  volume = {109},
  issue = {16},
  pages = {166805},
  numpages = {5},
  year = {2012},
  month = {Oct},
  publisher = {American Physical Society},
  doi = {10.1103/PhysRevLett.109.166805},
  url = {https://link.aps.org/doi/10.1103/PhysRevLett.109.166805}
}

@article{6,
   title = {Strong Spin-Orbit Splitting on Bi Surfaces},
  author = {Koroteev, Yu. M. and Bihlmayer, G. and Gayone, J. E. and Chulkov, E. V. and Bl\"ugel, S. and Echenique, P. M. and Hofmann, Ph.},
  journal = {Physics Review Letters},
  volume = {93},
  issue = {4},
  pages = {046403},
  numpages = {4},
  year = {2004},
  month = {Jul},
  publisher = {American Physical Society},
  doi = {10.1103/PhysRevLett.93.046403},
  url = {https://link.aps.org/doi/10.1103/PhysRevLett.93.046403}
}

@article{7,
author = {Makushko, Pavlo and Kovalev, Sergey and Zabila, Yevhen and Ilyakov, Igor and Ponomaryov, Alexey and Arshad, Atiqa and Prajapati, Gulloo and Oliveira, Thales and Deinert, Jan-Christoph and Chekhonin, Paul and Veremchuk, Igor and Kosub, Tobias and Skourski, Yurii and Ganss, Fabian and Makarov, Denys and Ortix, C.},
year = {2024},
month = {02},
pages = {1-9},
title = {A tunable room-temperature nonlinear Hall effect in elemental bismuth thin films},
volume = {7},
journal = {Nature Electronics},
doi = {10.1038/s41928-024-01118-y}
}

@inproceedings{7.2,
author = {F. Boivin and Oulin Yu and A. Silberztein and E. Y. J. Ma and G. Gervais},
title = {{Temperature-independent anomalous Hall effect and magnetic field angular dependence in thin bismuth electronic devices}},
volume = {13586},
booktitle = {Spintronics XVIII},
editor = {Jean-Eric Wegrowe and Joseph S. Friedman and Manijeh Razeghi and Henri Jaffr{\`e}s},
organization = {International Society for Optics and Photonics},
publisher = {SPIE},
pages = {1358608},
year = {2025},
doi = {10.1117/12.3064514},
URL = {https://doi.org/10.1117/12.3064514}
}

@article{8,
    title = {Observation of Temperature-Independent Anomalous Hall Effect in Thin Bismuth from Near Absolute Zero to 300 K Temperature},
  author = {Yu, Oulin and Boivin, F. and Silberztein, A. and Gervais, G.},
  journal = {Physics Review Letters},
  volume = {134},
  issue = {6},
  pages = {066603},
  numpages = {6},
  year = {2025},
  month = {Feb},
  publisher = {American Physical Society},
  doi = {10.1103/PhysRevLett.134.066603},
  url = {https://link.aps.org/doi/10.1103/PhysRevLett.134.066603}
}

@article{9,
title = {Higher-order topology in bismuth},
author = {Schindler, Frank and Wang, Zhijun and Vergniory, Maia G. and Cook, Ashley M. and Murani, Anil and Sengupta, Shamashis and Kasumov, Alik Yu. and Deblock, Richard and Jeon, Sangjun and Drozdov, Ilya and Bouchiat, Hélène and Guéron, Sophie and Yazdani, Ali and Bernevig, B. Andrei and Neupert, Titus},
doi = {10.1038/s41567-018-0224-7},
journal = {Nature Physics},
number = 9,
volume = 14,
place = {United States},
year = {2018},
month = {7}
}

@article{10,
title = {The surfaces of bismuth: Structural and electronic properties},
journal = {Progress in Surface Science},
volume = {81},
number = {5},
pages = {191-245},
year = {2006},
issn = {0079-6816},
doi = {https://doi.org/10.1016/j.progsurf.2006.03.001},
url = {https://www.sciencedirect.com/science/article/pii/S0079681606000232},
author = {Ph. Hofmann},
}

@article{11,
title = {Specific features of bismuth layers condensed at liquid helium temperature},
journal = {Doklady Akademii Nauk SSSR},
volume = {194},
number = {2},
pages = {301-305},
year = {1970},
url = {https://www.mathnet.ru/php/archive.phtml?wshow=paper&jrnid=dan&paperid=35671&option_lang=eng},
author = {Lazarev, B. G. and Kuz'menk and V. M. and Sudovtsov, A. I. and  Melnikov V. M. },
}

@article{12,
title = {Amorphous metals and their superconductivity},
journal = {Physics Reports},
volume = {27},
number = {4},
pages = {159-185},
year = {1976},
issn = {0370-1573},
doi = {https://doi.org/10.1016/0370-1573(76)90040-5},
url = {https://www.sciencedirect.com/science/article/pii/0370157376900405},
author = {G. Bergmann},
}

@article{13,
  title = {Spin-orbit interaction and phase coherence in lithographically defined bismuth wires},
  author = {Rudolph, M. and Heremans, J. J.},
  journal = {Physics Review B},
  volume = {83},
  issue = {20},
  pages = {205410},
  numpages = {6},
  year = {2011},
  month = {May},
  publisher = {American Physical Society},
  doi = {10.1103/PhysRevB.83.205410},
  url = {https://link.aps.org/doi/10.1103/PhysRevB.83.205410}
}

@article{14,
    author = {Rudolph, M. and Heremans, J. J.},
    title = {Electronic and quantum phase coherence properties of bismuth thin films},
    journal = {Applied Physics Letters},
    volume = {100},
    number = {24},
    pages = {241601},
    year = {2012},
    month = {06},
    issn = {0003-6951},
    doi = {10.1063/1.4729035},
    url = {https://doi.org/10.1063/1.4729035},
}

@article{15,
   author = {Song, Fei and Wells, Justin W. and Jiang, Zheng and Saxegaard, Magne and Wahlström, Erik},
   title = {Low-Temperature Growth of Bismuth Thin Films with (111) Facet on Highly Oriented Pyrolytic Graphite},
   journal = {ACS Applied Materials and Interfaces},
   volume = {7},
   number = {16},
   pages = {8525-8532},
   note = {doi: 10.1021/acsami.5b00264},
   ISSN = {1944-8244},
   DOI = {10.1021/acsami.5b00264},
   url = {https://doi.org/10.1021/acsami.5b00264},
   year = {2015},
   type = {Journal Article}
}

@article{16,
doi = {10.1143/JJAP.5.764},
url = {https://doi.org/10.1143/JJAP.5.764},
year = {1966},
month = {sep},
publisher = {},
volume = {5},
number = {9},
pages = {764},
author = {Fujime, Satoru},
title = {Electron Diffraction at Low Temperature II. Radial Distribution Analysis of Metastable Structure of Metal Films Prepa ed by Low Temperature Condensation},
journal = {Japanese Journal of Applied Physics},
}

@article{17,
author = {Om Prakash  and Anil Kumar  and A. Thamizhavel  and S. Ramakrishnan },
title = {Evidence for bulk superconductivity in pure bismuth single crystals at ambient pressure},
journal = {Science},
volume = {355},
number = {6320},
pages = {52-55},
year = {2017},
doi = {10.1126/science.aaf8227},
URL = {https://www.science.org/doi/abs/10.1126/science.aaf8227},
}

@article{18,
title = {A structure zone diagram including plasma-based deposition and ion etching},
journal = {Thin Solid Films},
volume = {518},
number = {15},
pages = {4087-4090},
year = {2010},
issn = {0040-6090},
doi = {https://doi.org/10.1016/j.tsf.2009.10.145},
url = {https://www.sciencedirect.com/science/article/pii/S0040609009018288},
author = {André Anders},
}

@article{19,
title = {Investigation of growth characteristics and semimetal–semiconductor transition of polycrystalline bis­muth thin films},
journal = {IUCrJ},
volume = {7},
pages = {49-57},
year = {2020},
issn = {2052-2525},
doi = {https://doi.org/10.1107/S2052252519015458},
url = {https://www.sciencedirect.com/science/article/pii/S2052252522008958},
author = {Nan Wang and Yu-Xiang Dai and Tian-Lin Wang and Hua-Zhe Yang and Yang Qi},
}

@article{20,
  title = {Nanofilm Allotrope and Phase Transformation of Ultrathin Bi Film on $\mathrm{S}\mathrm{i}(111)\mathrm{\text{\ensuremath{-}}}7\ifmmode\times\else\texttimes\fi{}7$},
  author = {Nagao, T. and Sadowski, J. T. and Saito, M. and Yaginuma, S. and Fujikawa, Y. and Kogure, T. and Ohno, T. and Hasegawa, Y. and Hasegawa, S. and Sakurai, T.},
  journal = {Physics Review Letters},
  volume = {93},
  issue = {10},
  pages = {105501},
  numpages = {4},
  year = {2004},
  month = {Aug},
  publisher = {American Physical Society},
  doi = {10.1103/PhysRevLett.93.105501},
  url = {https://link.aps.org/doi/10.1103/PhysRevLett.93.105501}
}

@article{21,
title = {Van der Waals epitaxy—a new epitaxial growth method for a highly lattice-mismatched system},
journal = {Thin Solid Films},
volume = {216},
number = {1},
pages = {72-76},
year = {1992},
note = {Papers presented at the International Workshop on Science and Technology of Thin Films for the 21st Century, Evanston, IL, USA, July 28-August 2, 1991},
issn = {0040-6090},
doi = {https://doi.org/10.1016/0040-6090(92)90872-9},
url = {https://www.sciencedirect.com/science/article/pii/0040609092908729},
author = {Atsushi Koma},
}

@article{22,
    author = {Galbiati, Miriam and Motta, Nunzio and De Crescenzi, Maurizio and Camilli, Luca},
    title = {Group-IV 2D materials beyond graphene on nonmetal substrates: Challenges, recent progress, and future perspectives},
    journal = {Applied Physics Reviews},
    volume = {6},
    number = {4},
    pages = {041310},
    year = {2019},
    month = {11},
    issn = {1931-9401},
    doi = {10.1063/1.5121276},
    url = {https://doi.org/10.1063/1.5121276},
}

@article{23,
author = {Littlejohn, Aaron and Xiang, Yu and Rauch, E and Lu, T.-M and Wang, G.-C},
year = {2017},
month = {11},
pages = {185305},
title = {Van der Waals epitaxy of Ge films on mica},
volume = {122},
journal = {Journal of Applied Physics},
doi = {10.1063/1.5000502}
}

@article{23.5,
   author = {Chen, Laisi and Wu, Amy X. and Tulu, Naol and Wang, Joshua and Juanson, Adrian and Watanabe, Kenji and Taniguchi, Takashi and Pettes, Michael T. and Campbell, Marshall A. and Xu, Mingjie and Gadre, Chaitanya A. and Zhou, Yinong and Chen, Hangman and Cao, Penghui and Jauregui, Luis A. and Wu, Ruqian and Pan, Xiaoqing and Sanchez-Yamagishi, Javier D.},
   title = {Exceptional electronic transport and quantum oscillations in thin bismuth crystals grown inside van der Waals materials},
   journal = {Nature Materials},
   volume = {23},
   number = {6},
   pages = {741-746},
   ISSN = {1476-4660},
   DOI = {10.1038/s41563-024-01894-0},
   url = {https://doi.org/10.1038/s41563-024-01894-0},
   year = {2024},
   type = {Journal Article}
}

@phdthesis{24,
    school = {Ecole Polytechnique Fédérale de Lausanne},
   author = {Beaulieu, Guillaume},
   title = {Preparetion of Amorphous Bismuth Films by Quench Condensation Inside of a Superconducting Magnet},
   university = { École Polytechnique Fédérale de Lausanne},
   year = {2022},
   type = {Thesis}
}

@article{25,
  title = {Carrier properties of Bi(111) grown on mica and Si(111)},
  author = {Jiang, Zijian and Soghomonian, V. and Heremans, J. J.},
  journal = {Physics Review Materials},
  volume = {6},
  issue = {9},
  pages = {095003},
  numpages = {8},
  year = {2022},
  month = {Sep},
  publisher = {American Physical Society},
  doi = {10.1103/PhysRevMaterials.6.095003},
  url = {https://link.aps.org/doi/10.1103/PhysRevMaterials.6.095003}
}

@article{26,
title = {Theory and experiments in epitaxial growth},
journal = {Materials Chemistry and Physics},
volume = {9},
number = {1},
pages = {93-116},
year = {1983},
issn = {0254-0584},
doi = {https://doi.org/10.1016/0254-0584(82)90012-8},
url = {https://www.sciencedirect.com/science/article/pii/0254058482900128},
author = {I. Markov},
}

@article{27,
   author = {Jankowski, M. and Kamiński, D. and Vergeer, K. and Mirolo, M. and Carla, F. and Rijnders, G. and Bollmann, T. R.},
   title = {Controlling the growth of Bi(110) and Bi(111) films on an insulating substrate},
   journal = {Nanotechnology},
   volume = {28},
   number = {15},
   pages = {155602},
   ISSN = {0957-4484},
   DOI = {10.1088/1361-6528/aa61dd},
   year = {2017},
   type = {Journal Article}
}

@book{28,
   author = {Dubrovskii, Vladimir G.},
   title = {Nucleation Theory and Growth of Nanostructures},
   year = {2016},
   type = {Book},
   publisher = {Springer}
}

@article{29,
  title = {Random-deposition models for thin-film epitaxial growth},
  author = {Evans, J. W.},
  journal = {Physics Review B},
  volume = {39},
  issue = {9},
  pages = {5655--5664},
  numpages = {0},
  year = {1989},
  month = {Mar},
  publisher = {American Physical Society},
  doi = {10.1103/PhysRevB.39.5655},
  url = {https://link.aps.org/doi/10.1103/PhysRevB.39.5655}
}

@article{30,
  title = {The Scherrer Formula for X-Ray Particle Size Determination},
  author = {Patterson, A. L.},
  journal = {Physics Review},
  volume = {56},
  issue = {10},
  pages = {978--982},
  numpages = {0},
  year = {1939},
  month = {Nov},
  publisher = {American Physical Society},
  doi = {10.1103/PhysRev.56.978},
  url = {https://link.aps.org/doi/10.1103/PhysRev.56.978}
}

@article{31,
  title = {Role of the surface states in the magnetotransport properties of ultrathin bismuth films},
  author = {Marcano, N. and Sangiao, S. and Mag\'en, C. and Morell\'on, L. and Ibarra, M. R. and Plaza, M. and P\'erez, L. and De Teresa, J. M.},
  journal = {Physics Review B},
  volume = {82},
  issue = {12},
  pages = {125326},
  numpages = {6},
  year = {2010},
  month = {Sep},
  publisher = {American Physical Society},
  doi = {10.1103/PhysRevB.82.125326},
  url = {https://link.aps.org/doi/10.1103/PhysRevB.82.125326}
}

@article{32,
author = {Yang, Zhibin and Wu, Zehan and Lyu, Yongxin and Hao, Jianhua},
title = {Centimeter-scale growth of two-dimensional layered high-mobility bismuth films by pulsed laser deposition},
journal = {InfoMat},
volume = {1},
number = {1},
pages = {98-107},
doi = {https://doi.org/10.1002/inf2.12001},
url = {https://onlinelibrary.wiley.com/doi/abs/10.1002/inf2.12001},
year = {2019}
}

@article{33,
    author = {Arisaka, Taichi and Otsuka, Mioko and Hasegawa, Yasuhiro},
    title = {Investigation of carrier scattering process in polycrystalline bulk bismuth at 300 K},
    journal = {Journal of Applied Physics},
    volume = {123},
    number = {23},
    pages = {235107},
    year = {2018},
    month = {06},
    issn = {0021-8979},
    doi = {10.1063/1.5032137},
    url = {https://doi.org/10.1063/1.5032137},
}

@article{34,
title = {Strain-engineered conductivity transition and its mechanism in Bi (110) Film},
journal = {Surfaces and Interfaces},
volume = {56},
pages = {105595},
year = {2025},
issn = {2468-0230},
doi = {https://doi.org/10.1016/j.surfin.2024.105595},
url = {https://www.sciencedirect.com/science/article/pii/S2468023024017504},
author = {Y.L. Liu and X.H. Ma and W.P. Jia and S.H. Ren and H. Zhang and Y. Qi},
}

@article{35,
   author = {Yang, F. Y. and Liu, K. and Hong, K. and Reich, D. H. and Searson, P. C. and Chien, C. L.},
   title = {Large magnetoresistance of electrodeposited single-crystal bismuth thin films},
   journal = {Science},
   volume = {284},
   number = {5418},
   pages = {1335-7},
   ISSN = {0036-8075},
   DOI = {10.1126/science.284.5418.1335},
   year = {1999},
   type = {Journal Article}
}

@article{36,
   author = {Sandomirskii, V. B.},
   title = {Quantum Size Effects in a Semimetal Film},
   journal = {Soviet Physics JETP},
   volume = {52},
   year = {1967},
   type = {Journal Article},
   url = {https://jetp.ras.ru/cgi-bin/dn/e_025_01_0101.pdf}
}

@article{37,
title = {Quantum size effect and electric conductivity in thin films of pure bismuth},
journal = {Journal of Physics and Chemistry of Solids},
volume = {53},
number = {8},
pages = {1059-1065},
year = {1992},
issn = {0022-3697},
doi = {https://doi.org/10.1016/0022-3697(92)90078-R},
url = {https://www.sciencedirect.com/science/article/pii/002236979290078R},
author = {H.T. Chu and W. Zhang},
}

@article{38,
  title = {Quantum transport in the surface states of epitaxial Bi(111) thin films},
  author = {Zhu, Kai and Wu, Lin and Gong, Xinxin and Xiao, Shunhao and Jin, Xiaofeng},
  journal = {Physics Review B},
  volume = {94},
  issue = {12},
  pages = {121401},
  numpages = {6},
  year = {2016},
  month = {Sep},
  publisher = {American Physical Society},
  doi = {10.1103/PhysRevB.94.121401},
  url = {https://link.aps.org/doi/10.1103/PhysRevB.94.121401}
}

\end{document}